\documentclass{article}
\usepackage[T1]{fontenc} 
\usepackage[utf8]{inputenc} 
\usepackage{ismir,amsmath,cite,url}
\usepackage{graphicx}
\usepackage{color}
\usepackage{amsfonts}
\usepackage{booktabs}

\usepackage[linesnumbered,ruled,vlined]{algorithm2e}
\RestyleAlgo{ruled}

\newcommand{\unrolled}{logical}

\newcommand{\mdiff}{\texttt{MDiff}}
\newcommand{\merr}{\texttt{MErr}}
\newcommand{\mstd}{\texttt{MDev}}
\newcommand{\macc}{\texttt{MAcc}}

\newcommand{\mbb}{\mathbb{M}}
\newcommand{\mbbstar}{\mbb^*}
\newcommand{\mbbest}{\mbb'}
\newcommand{\bb}{b}
\newcommand{\bcurr}{\bb_{\mindex}}

\newcommand{\floorindex}[1]{\left \lfloor #1 \right \rfloor}
\newcommand{\mindex}{\floorindex{m}}
\newcommand{\mfrac}{m - \mindex}

\usepackage{xcolor}
\usepackage{color-edits}
\addauthor[Michael]{mf}{purple}
\addauthor[Irmak]{ib}{blue}

\title{Just label the repeats \\ for in-the-wild audio-to-score alignment}

\threeauthors
  {Irmak Bukey} {Carnegie Mellon University}
  {Michael Feffer} {Carnegie Mellon University}
  {Chris Donahue} {Carnegie Mellon University}

\def\authorname{I. Bukey, M. Feffer, and C. Donahue}

\usepackage[bookmarks=false,pdfauthor={\authorname},pdfsubject={\papersubject},hidelinks]{hyperref}

\usepackage{cleveref}

\begin{document}

\maketitle

\begin{abstract}
We propose an efficient workflow for high-quality offline alignment of in-the-wild performance audio and corresponding sheet music scans (images).\footnote{Video examples: \url{https://bit.ly/jltr-ismir2024}\\Code: \url{https://github.com/irmakbky/jltr-alignment}\\Corresponding author: Irmak Bukey <\href{mailto:ibukey@cs.cmu.edu}{ibukey@cs.cmu.edu}>}
Recent work on audio-to-score alignment extends dynamic time warping~(DTW) to be theoretically able to handle \emph{jumps} in sheet music induced by repeat signs---this method requires no human annotations, but we show that it often yields low-quality alignments. 
As an alternative, 
we propose a workflow and interface that allows users to quickly annotate jumps (by clicking on repeat signs), 
requiring a small amount of human supervision but yielding much higher quality alignments on average.
Additionally, we refine audio and score feature representations to improve alignment quality by:
(1)~integrating measure detection into the score feature representation, and
(2)~using raw onset prediction probabilities from a music transcription model instead of piano roll. 
We propose an evaluation protocol for audio-to-score alignment that computes the distance between the estimated and ground truth alignment in units of measures. 
Under this evaluation, we find that our proposed jump annotation workflow and improved feature representations together improve alignment accuracy by 
$150\%$ 
relative to prior work $(33\% \to 82\%$).
\vspace{-2.25mm}
\end{abstract}
\begin{figure*}[t]
    \centering
    \includegraphics[alt={A figure consisting of two subfigures. The first is a diagram illustrating the task of audio-to-score alignment. It shows that the inputs into the system consist of a musical score, repeats, and audio, while the output shows the alignment of the score with numbered sections indicating the order of playback. The second subfigure is a flowchart of our proposed system for audio-to-score alignment. Inputs include score (PDF), repeat labels, and audio (MP3). The system works by passing the score PDF into measure detection which returns bounding boxes that then get used in producing a bootleg score. The audio is transcribed to get note onset probabilities. The bootleg scores and note onset probabilities are then aligned using the DTW algorithm, producing the final alignment.},width=\textwidth]{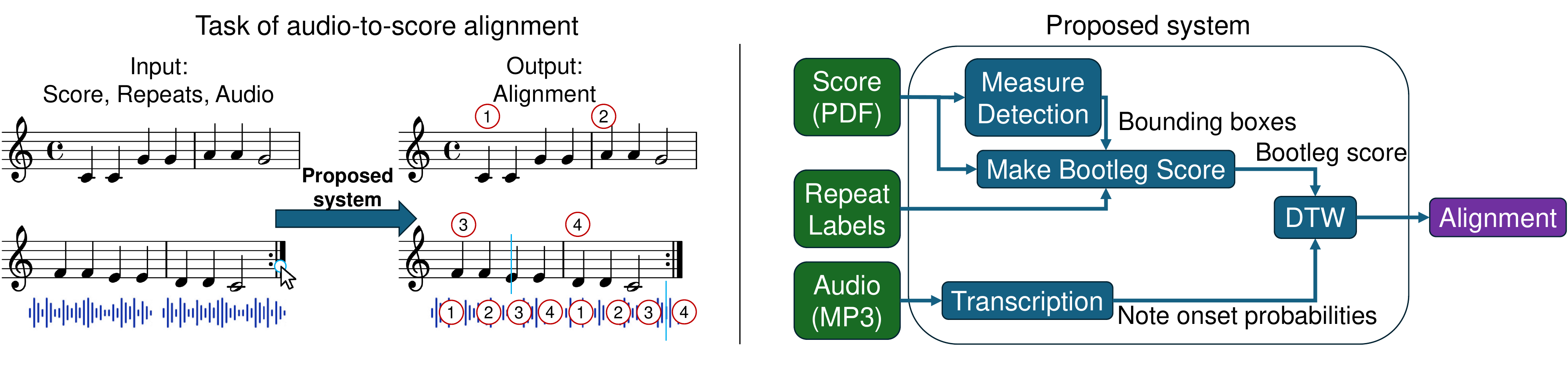}
    \caption{An overview of the task of audio-to-score alignment and our proposed approach. 
    Given a score image (as a PDF) and corresponding performance audio (e.g.,~an MP3) as input, the task involves outputting an alignment between time in the recording and playheads in the score image. 
    A key challenge in this task is handling jumps in the score, e.g., those created by repeat signs. 
    In lieu of robust automatic methods for detecting or handling jumps, we propose a pragmatic approach of having experts simply label the repeats, which can be done quickly and greatly improves task performance. 
    Our proposed system combines the repeat labels with score feature representations inspired by past work on bootleg scores~\cite{yang2020midi}.
    This score representation is aligned with audio feature representations inspired by~\cite{maman2022unaligned} using ordinary DTW.
    }
    \label{fig:overview}
\end{figure*}

\section{Introduction}\label{sec:introduction}


Sheet music has been used as a primary means of communicating musical ideas for centuries. 
Accordingly, sheet music is a profoundly important modality for MIR, 
not only because of the breadth of musical knowledge and history contained within, 
but also because sheet music constitutes a vital interface between MIR systems and musicians. 
However, while multimodal MIR systems are rapidly improving at tasks like music transcription~\cite{hawthorne2017onsets,donahue2022melody,gardner2021mt3,bittner2022basicpitch} 
and controllable generation~\cite{agostinelli2023musiclm,copet2024simple,wu2024music}, 
these systems typically operate on MIDI as a symbolic music format. 
This may be less useful to musicians, e.g.,~a musician might prefer transcription systems to output sheet music instead of MIDI. 


We conjecture that the scarcity of fine-grained alignment data linking sheet music to corresponding performance audio is a key bottleneck to incorporating sheet music into multimodal MIR systems. 
Alignments allow multimodal MIR data to be segmented into input-output chunks of tractable length for training models, 
and the lack of sheet music alignments may partially explain why sheet music is mostly overlooked. 
Moreover, alignments have practical utility outside of multimodal MIR, e.g.,~they may be used by musicians to practice along with pre-recorded accompaniments. 
Unfortunately, collecting alignments is deceptively tricky. 
For example, one could have a musician use a touch screen to point to the current location in sheet music while listening to a recording in real time. 
However, their tracking may be imprecise (due to expressive performance timing) 
and lack non-obvious details that are essential for segmentation (bar line locations, number of active staves). 


In this work, we investigate the task of alignment of offline in-the-wild performance audio and corresponding sheet music scans (images), with a long-term goal of aligning large corpora of sheet music and performance recordings at scale. 
Much of the past work on audio-to-score alignment make at least one of several common assumptions that inhibit their practicality for collecting aligned data at scale: 
(i)~the presumed availability of digital scores like MIDI or MusicXML as opposed to sheet music images~\cite{carabias2015audio,syue2017accurate,arzt2018audio,tanprasert2020midi,foscarin2020asap}, 
(ii)~the alignment of MIDI performances or synthesized audio instead of real audio recordings~\cite{dorfer2017learning,arzt2018audio,shan2020improved},
(iii)~limitations in instrument diversity, commonly piano only~\cite{carabias2015audio,shan2020improved,shan2021automatic}, or
(iv)~dependence on time-consuming human annotation~\cite{feffer2022assistive,Soundslice}. 


Here we propose an audio-to-score alignment procedure that makes none of these assumptions, potentially offering a path forward for large-scale data collection. 
Most closely related to our approach is that of Shan et al.~\cite{shan2020improved,shan2021automatic}, 
who examine offline alignment of in-the-wild piano sheet music images and performance recordings by aligning feature representations derived from the score and audio via MIR methods. 
In addition to operating on more diverse ensembles, 
our work has two primary distinctions: 
(1)~we take a different approach to handling \emph{jumps} in scores, and
(2)~we modify their feature representations. 


A key challenge in audio-to-score alignment is handling inter-measure jumps in scores induced by repeat signs. 
Shan et al.~\cite{shan2020improved,shan2021automatic} propose extensions to DTW that are capable of automatically handling jumps. 
Here we propose a pragmatic alternative: a workflow and interface that allows humans to quickly annotate jumps, and a system that incorporates these jump labels. 
We find that this approach can yield much higher-quality alignments than the automatic one, costing only seconds of annotator time.


We additionally extend the \emph{bootleg score} feature representations used by Shan et al. \cite{shan2021automatic}, first proposed by Yang et al. ~\cite{yang2020midi}. 
Creating a bootleg score involves detecting noteheads and staff lines to produce a simple binary representation of a score that is conducive to alignment. 
We find that the use of measure bounding box detection as a preprocessing step improves the quality of underlying notehead and staff line detection algorithms. 
Additionally, motivated by findings in~\cite{maman2022unaligned}, we find that using raw onset probabilities predicted by a music transcription model as the audio feature representation produces higher quality alignments than using the MIDI transcriptions---see~\Cref{fig:overview} for a summary.

Motivated by our long-term goals of bringing sheet music into multimodal MIR, 
we also propose a new measure-aware evaluation scheme for comparing alignments. 
We speculate that measure-level alignment granularity is necessary for tractable training of multimodal MIR systems in the short term and that human perception of alignment quality is tied to measures. 
Accordingly, 
we prescribe new measure-aware alignment metrics for this task, such as an accuracy metric which reports the proportion of time where the estimated alignment is within a half measure radius of the ground truth alignment. 
On a small but diverse dataset of in-the-wild sheet music and aligned audio~\cite{feffer2022assistive}, we observe that our proposed system achieves an accuracy of 
$120\%$ relative to that of Shan et al. ($33\% \to 72\%$). 
By providing repeat labels, we improve the absolute accuracy of our system from $20\% \to 83\%$ on a subset of pieces that have repeats.
Our work makes the following contributions:
\begin{itemize}
    \item A system capable of high-quality in-the-wild alignment of sheet music images and performance audio.
    \item A pragmatic workflow we call \emph{Just Label The Repeats} that further improves alignment accuracy.
    \item An interface that enables rapid jump annotation.
\end{itemize}

\vspace{-1.75mm}
\section{Task description}
\label{sec:task-desc}

\begin{figure}[ht]
    \centering
    \includegraphics[alt={Illustration of a score playhead in a score. There are two measures visible with corresponding bounding boxes, a dashed vertical line representing the estimated playhead in the score, and a solid vertical line representing the groundtruth position of the playhead. There are also lines indicating the horizontal offset, the vertical offset, and the height of the measure playhead, all defined relative to the page.},width=0.98\columnwidth]{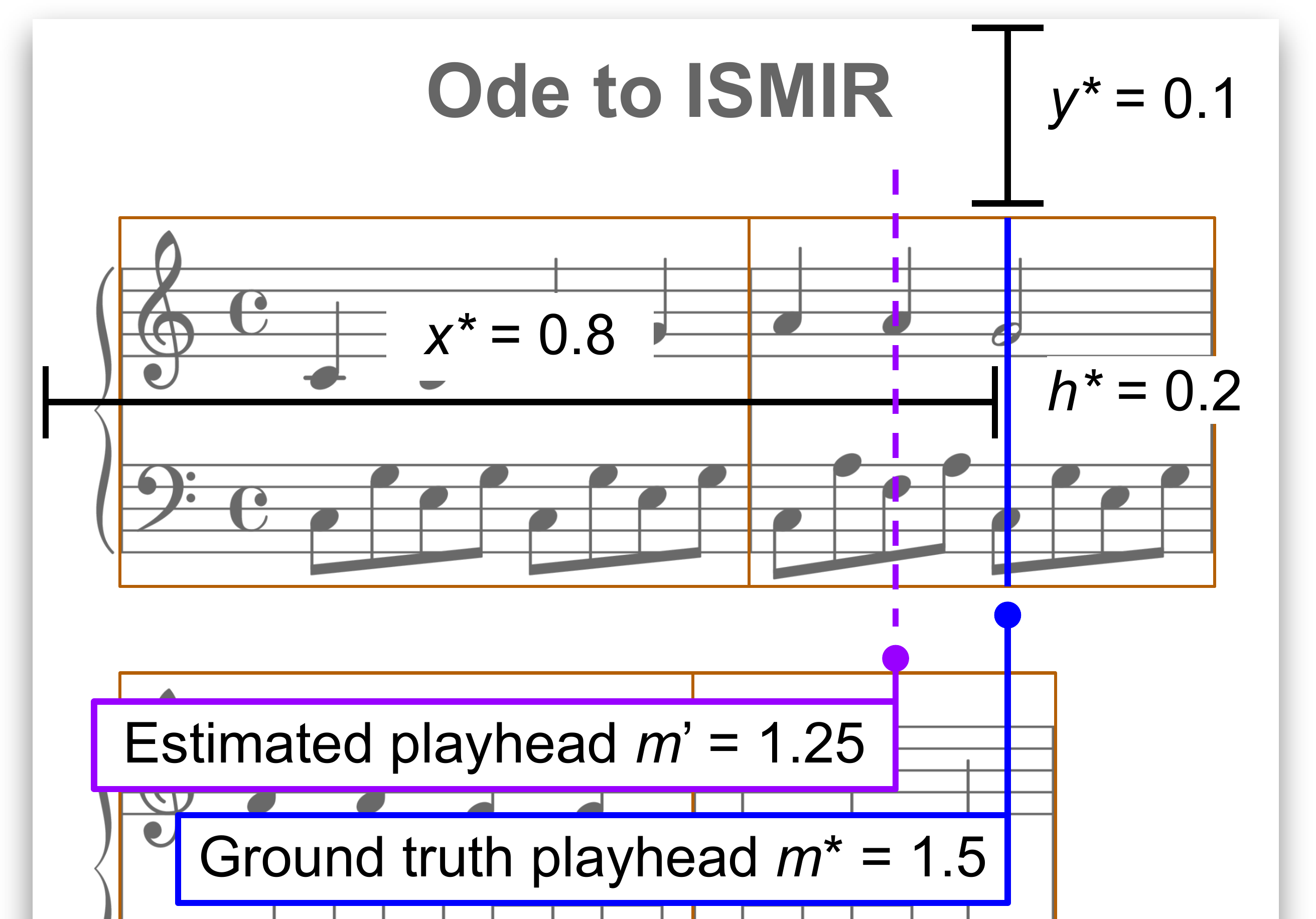}
    \caption{A score playhead (blue line), the output of an audio-to-score alignment, is characterized by its 
    vertical offset~($y$),
    horizontal offset~($x$),
    and height ($h$), all
    relative to the page. A measure-aware alignment is indexed by $m$, a fractional measure, that can be converted to a score playhead by lookup and interpolation in a list of bounding boxes (brown outlines). Our measure-aware evaluation compares estimated playheads $m'$ to ground truth $m^*$.}
    \label{fig:playhead}
\end{figure}

Motivated by 
Thickstun et al.~\cite{thickstun2020rethinking}, 
here we formalize both the task of in-the-wild audio-to-score alignment 
and our proposed measure-aware evaluation. 
For a sheet music image with $P$ pages (henceforth, a \emph{score}), we define a \emph{score playhead} (aligned position marker) as a tuple ${(p, y, h, x) \in \mathcal{S} = \{0, \ldots, P-1\} \times [0, 1]^3}$, 
where $p$ is the page number, 
$y$ and $h$ are the offset and height of the current \emph{system} (collection of staves) relative to the top edge and height of the page, and 
$x$ is the playhead offset 
relative to the left edge of the page (see~\Cref{fig:playhead}).
An analogous 
\emph{audio playhead} is comparatively straightforward: a timestamp ${t \in [0, T)}$, where $T$ is the audio length in seconds.
An \emph{alignment} is a mapping from audio to score playheads, i.e., 
${[0, T) \to \mathcal{S}}$.

\vspace{-2.25mm}
\subsection{Measure-aware alignment}
\label{sec:task_measure_aware}

The above task definition is intended to be broad enough to encompass both past and future work on this task. 
Here we define a more specific form of alignment based around the location and ordering of \emph{measures} in the score.

In this setting, measures are characterized by an ordered list of $M$ bounding boxes ${\mbb = [\bb_0, \ldots, \bb_{M-1}]}$, where ${\bb_i = (\bb^p_i, \bb^y_i, \bb^h_i, \bb^x_i, \bb^w_i)}$. 
Respectively, this tuple defines for each bounding box its
page number, vertical offset, height, horizontal offset, and width. 
The ordering of this list is defined as the \emph{\unrolled} order that an expert would traverse when performing the piece---all \emph{jumps} (repeat signs, Dal segno, etc.) are unrolled. 
For example, a score with $4$ measures and a repeat implies that ${M = 8}$ and ${\bb_i = \bb_{i+4}}$. 

Given a list of bounding boxes, 
a \emph{measure-aware} score playhead can be characterized by a single continuous value ${m \in [0, M)}$, where $\bcurr$ is the bounding box of the current measure and fractional residual $\mfrac$ represents the offset from the left edge of the bounding box relative to its width. 
To convert a measure-aware score playhead in $[0, M)$ to an ordinary score playhead in $\mathcal{S}$, 
we define ${h_{\mbb} : m \mapsto (\bcurr^p, \bcurr^y, \bcurr^h, \bcurr^x + \bcurr^w (\mfrac))}$.

Given $\mbb$, a \emph{measure-aware alignment} is a function ${g : [0, T) \to [0, M)}$. 
Because outputs of $g$ 
index  
\unrolled{} order (where jumps are unrolled), 
a measure-aware alignment $g$ is 
a monotonically increasing function, i.e., ${g(t_a) \leq g(t_b) \iff t_a \leq t_b}$. 
Furthermore, 
we can compose $h_{\mbb}$ and $g$ to induce an alignment that outputs score playheads, i.e.,~${h_{\mbb} \circ g : [0, T) \to \mathcal{S}}$.
\vspace{-2.25mm}
\subsection{Measure-aware evaluation}
\label{sec:measure_aware_eval}

Here we propose three measure-aware metrics for evaluating estimated alignments. 
\textbf{Our primary evaluation metric, $\macc$, is defined as the proportion of time where the estimated score playhead is within a half measure radius of the ground truth score playhead}, which we posit is sufficiently precise for broader goals of multimodal MIR systems. 
$\merr$ and $\mstd$ are the mean and standard deviation (across time) of the 
absolute error 
between the estimated and ground truth playheads in units of ground truth measures.

More formally, given a ground truth measure-aware alignment $g^*$ characterized by measures $\mbbstar$, and an estimated alignment $g'$ characterized by $\mbbest$, 
we define:
\begin{align*}
    \mdiff(t) &= \texttt{Reindex}(g'(t), \mbbest, \mbbstar) - g^{*}(t), \\
    \macc &\equiv \frac{1}{N} \sum_{i=0}^{N-1} 
        \begin{cases}
        1 & \text{if}~|\mdiff(\frac{Ti}{N})| \leq \frac{1}{2} \\
        0 & \text{otherwise}.
        \end{cases} \\
    \merr &\equiv \frac{1}{N} \sum_{i=0}^{N-1} \left| \mdiff \left( \frac{Ti}{N} \right) \right|, \\
    \mstd &\equiv \sqrt{\frac{1}{N} \sum_{i=0}^{N-1} \mdiff \left( \frac{Ti}{N} \right)^2}
\end{align*}
We set $N=100T$, i.e., we compute all metrics at a resolution of $100$ comparisons per second. 
As an example, \Cref{fig:playhead} shows a single comparison where $\mdiff=0.25$.

The \texttt{Reindex} procedure 
ensures that the evaluation is based in units of ground truth measure indices, 
despite potential discrepancies between $\mbbest$ and $\mbbstar$. 
Informally, 
this procedure matches each box in $\mbbest$ to the box in $\mbbstar$ that it is closest to in terms of Euclidean distance between midpoints, 
using the ordering from $\mbbstar$ to break ties that occur for repeated measures. 


\vspace{-2mm}
\vspace{-1.75mm}
\section{System description}
\vspace{-1mm}
In this section, we 
detail our proposed system for 
in-the-wild audio-to-score alignment. At a high level, our method 
uses DTW to 
align 
piano roll-like feature representations extracted independently from the audio and score 
(\Cref{fig:overview}).
Both representations are matrices where 
one axis corresponds to time (either in units of seconds or measures) and 
the other corresponds to $88$ MIDI pitches from A0~to~C8 (piano range).
Specifically, score feature representations are matrices in ${\{0,1\}^{48M \times 88}}$ (binary) where $M$ is the number of measures in the score, and audio feature representations are matrices in ${[0, 1]^{Tf_k \times 88}}$ (continuous) 
for performance length $T$ and frame rate $f_k = 31$Hz. 

\vspace{-2.25mm}
\subsection{Score feature representation}
\label{sec:score-feature-rep}

Our proposed score feature representation is an extension of \emph{bootleg scores} proposed by Yang et al.~\cite{yang2020midi}. 
Extracting bootleg scores involves detecting noteheads and staff lines, and then combining this information into a binary matrix where a $1$ encodes the presence of a notehead at a particular horizontal position on a particular staff line. 
Our system extends this representation in two ways:
(1)~we use measure detection as a preprocessing step and run notehead detection algorithms on segmented measure images instead of full pages, and
(2)~we translate notehead positions on staff lines into MIDI pitches before alignment.

\textbf{Measure, notehead and staff line detection.} 
We use the methods from \cite{yang2020midi} to detect noteheads and staff lines from score images.
Instead of operating on full page images, we first segment pages into measures using the measure detection model from~\cite{waloschek2019identification}, 
and detect noteheads and staff lines on individual measure images resized to fixed dimensions. 
In preliminary experiments, we found that measure segmentation improved detection consistency---our intuition is that these methods are sensitive to absolute pixel sizes of noteheads and stafflines, and resizing individual measures to uniform size reduces variance in the sizes of these attributes across measures and pieces. 
Here we resize measures by resizing the smaller of their height and width to $900$ pixels, preserving aspect ratio for the larger of the two.
For each notehead, we retain its \emph{bootleg location}, defined as its raw (pixel-wise) horizontal location within the measure, and its semantic (discrete) vertical position within the detected staff lines (e.g.,~an F and G natural in treble clef are one apart in their \emph{staff positions}).

\textbf{Piano rolls.} 
We create a binary piano roll-like representation of the score using the bootleg locations of noteheads.
Specifically, 
for each \unrolled{} measure image index ${k \in \{0, \ldots, M - 1\}}$, 
we construct a binary matrix $S_k \in \{0,1\}^{48 \times 88}$,  
where a $1$ at row $i$ and column $j$ corresponds to a notehead with horizontal location $\frac{i}{48}$ relative to the measure width, and its staff position converted to a pitch $j$. We pick a measure representation to have 48 rows to give sufficient resolution to a variety of rhythmic patterns and note durations.
We concatenate $[S_0, \ldots, S_{M-1}]$ together to form our final representation $S \in \{0,1\}^{48M \times 88}$.

\textbf{Converting staff position to pitch.} A key obstacle is that, without key signature and clef information, the mapping 
from staff positions to MIDI pitches is ambiguous. 
If we had these attributes, we could simply look up the pitch associated with each staff line. 
However, 
we found existing OMR systems to have brittle support for detecting this information for in-the-wild sheet music---hence, 
our method does not assume that we have access to this information. 
Accordingly, we convert bootleg scores into piano rolls by simply assuming treble and bass clefs respectively when two staves are detected---if more or fewer staves are detected, we default to the treble clef--- and the key of C major. 
Surprisingly, 
perhaps because of the global optimality of DTW, 
these assumptions lead to reasonable alignments even when they are incorrect.
We note that ground truth key signature and clef information for each measure can be fed into our method to facilitate and improve this conversion, but we do not require it.

\vspace{-2.25mm}
\subsection{Audio feature representation}
\label{sec:system-audio-feature-rep}

Our audio feature representation pipeline is comparatively simple. 
To compute it, we simply pass the audio through the Onsets and Frames piano transcription model~\cite{hawthorne2017onsets}. Motivated by \cite{maman2022unaligned}, we use the raw onset prediction probabilities from this model as our audio feature representation, which is a matrix in $[0, 1]^{Tf_k \times 88}$.
Despite this transcription model being trained on piano, we find that its onset probabilities can yield reasonable alignments even for non-piano audio.

\vspace{-2.25mm}
\subsection{Alignment}

Finally, we align the score representations in $\{0,1\}^{48M\times 88}$ and audio representations in $\mathbb{R}^{Tf_k \times 88}$.
We use the implementation of standard DTW from \texttt{librosa}~\cite{mcfee2015librosa} with default parameters: equal-weighted transitions $(1,~1)$, $(0,~1)$, and $(1,~0)$, and Euclidean distance to compute costs.



\vspace{-1.75mm}
\section{Experiments}


Here we detail our experiments, which center around comparing our proposed method to that of Shan et al.~\cite{shan2021automatic} on the MeSA-13~\cite{feffer2022assistive} and SMR~\cite{yang2020midi} datasets using our proposed measure-level evaluation (see~\Cref{sec:measure_aware_eval}).
\vspace{-2.25mm}
\subsection{Datasets}

We evaluate our approach and relevant baselines on two different datasets. 
The first is MeSA-13~\cite{feffer2022assistive}, a dataset of $13$ 
sheet music scans and corresponding real (i.e.,~not synthesized) performance audio. 
This dataset contains expert annotations of measure bounding boxes in logical order ($\mbbstar$) and the timestamp of every measure in the performance audio (we linearly interpolate between timestamps to get a continuous ground truth mapping $g^*$). 
While small, 
MeSA-13 has reasonable diversity in score typesetting, performance acoustics (two pieces feature instruments besides piano), and jumps (two pieces have repeats). 

The second dataset is a subset of the Sheet MIDI Retrieval v1.0 (SMR) dataset~\cite{yang2020midi}. 
The full dataset contains 
scanned scores from IMSLP for $100$ solo piano pieces (none of which have jumps), 
corresponding MIDI performances synthesized as audio, 
and 
human annotations of measures per line and measure timestamps. 
Of notable absence are annotations of measure bounding boxes, which are required for our proposed evaluation. 
Accordingly, 
we detect measures~\cite{waloschek2019identification} and discard pieces where the detections do not agree with annotations of measures per line---this leaves us with a subset of $60$ pieces for evaluation. 
Henceforth, SMR refers to this subset.



\vspace{-2.25mm}
\subsection{Access to additional annotations}
\label{sec:extra}

We primarily evaluate systems in an automatic setting where systems are only given the score and audio as input. 
Because our system can incorporate additional score annotations when available, we also evaluate in settings where our system has access to additional annotations from the ground truth, 
simulating workflows where experts are in the loop during alignment. 
Specifically, we explore settings where our method has access to ground truth repeat annotations (R), measure bounding boxes (M), and staff information (S)---clef and key signatures. 
We only evaluate in these settings on MeSA-13 where we have these labels.


\vspace{-2.25mm}
\subsection{Baselines}

We compare the performance of our system, composed of our feature extraction pipeline and vanilla DTW, to that of \cite{shan2021automatic}. In the latter system, the feature extraction pipeline uses bootleg scores and staff line detection on the score images to extract staff lines (referred to as segments). For audio features, a transcribed MIDI representation is obtained from the Onsets and Frames piano transcription model \cite{hawthorne2017onsets} which is then used to compute bootleg scores. Finally, Hierarchical DTW performs a segment-level alignment between score features and audio features while handling jumps and repeats that occur at segment boundaries, but not those within segments.
Thus, we opt to evaluate this baseline approach at the measure level instead of the segment level (which also allows for comparison with our system) by converting the segment-level alignment to a measure-level one via an algorithm with several key steps.

We first use measure detection \cite{waloschek2019identification} 
to locate measures in each segment. 
Then, we map segment indices to measure indices based on the positions of detected measures.
Finally, we turn the given alignment between audio timestamps and segment indices to one between audio timestamps and measure indices via linear interpolation.

We compare these two systems as proposed instead of 
comparing their components 
for two reasons. First, in \cite{shan2021automatic} it is claimed the system performs segment-level alignment on pieces without repeat info; we aim to test this. Second, while Hierarchical DTW allows backwards jumps to prior segments, it only allows forward jumps to one segment past the last one seen. Using Hierarchical DTW at the measure level would limit possible forward jumps to only one measure past the last one observed, which is insufficient for realistic alignment tasks. 


\subsection{Results and discussion}

\newcommand{\mathbest}[1]{#1}

\begin{table}[t]
 \begin{center}
 \begin{tabular}{lllccc}
\toprule
Dataset	& Given & System & \macc	& \merr	& \mstd \\
\midrule
M13
        & - & \cite{shan2021automatic} & 0.33 & 10.9 & 11.6	 \\
        & - & Ours	&  0.72 & 1.9 & 3.7 \\
\cmidrule{2-6}
        & R\textsuperscript{\textdagger} & Ours	& 0.82 & 0.4 & 0.2 \\
	& R,M & Ours	& 0.86 & 0.4 & 0.2\\
	& R,M,S & Ours	& 0.88 & 0.3 & 0.2  \\
\midrule 

$\text{M13}_{\text{R}}$
	& - &  \cite{shan2021automatic}	& 0.17 & 23.6 & 10.2\\
	& - & Ours	&0.20 & 10.0 & 3.0\\
\cmidrule{2-6}
        & R\textsuperscript{\textdagger} & Ours & 0.83 & 0.3 & 0.0\\
	& R,M & Ours	& 0.93 & 0.2 & 0.0\\
	& R,M,S & Ours	& 0.95 & 0.2 & 0.0\\
\midrule 

$\text{M13}_{\text{NR}}$
	& - & \cite{shan2021automatic}	& 0.36 & 8.6 & 10.2\\
	& - & Ours	&0.82 & 0.4 & 0.2\\
\cmidrule{2-6}
        & R\textsuperscript{\textdagger}	& Ours & 0.82 & 0.4 & 0.2\\
	& R,M & Ours	&0.85 & 0.4 & 0.3\\
	& R,M,S & Ours	&0.87 & 0.3 & 0.2\\
 \midrule
SMR
	& - & \cite{shan2021automatic} & 0.36 & 14.2 & 18.7\\
        & - & Ours	& 0.82 & 2.9 & 18.8\\
\bottomrule
 \end{tabular}
\end{center}
\caption{Evaluation on MeSA-13 (including subsets with \underline{R}epeats and \underline{N}o \underline{R}epeats) and SMR. 
 Our method outperforms that of~\cite{shan2021automatic} across all datasets except the subset of MeSA-13 with no repeats. 
 R\textsuperscript{\textdagger} is our recommended setting where our method is given access to ground truth \underline{R}epeats that require little time for humans to annotate---we observe limited gains from more time-consuming annotations of \underline{M}easure bounding boxes and \underline{S}taff metadata.}
 \label{tab:eval_all}
\end{table}


In Table \ref{tab:eval_all}, we report the measure-level alignment metrics (\Cref{sec:measure_aware_eval}) of our system
in all four settings and the system of~\cite{shan2021automatic} 
across the MeSA-13 (M13) and SMR datasets.
To emphasize the effect of jumps on alignment performance, we separately report performance on the subset of MeSA-13 pieces with and without repeats ($\text{M13}_{\text{R}}$ and $\text{M13}_{\text{NR}}$, respectively).


We observe that our system outperforms that of \cite{shan2021automatic} in the automatic setting
across all datasets.
The superior performance of our system over that of \cite{shan2021automatic} is likely due to our refinements to feature representations. We also note here that the system of \cite{shan2021automatic} is designed to work on line-level, therefore evaluating it using our measure-level metric yields lower accuracy than what was reported in \cite{shan2021automatic}.

Additionally, we observe that given repeats and ground truth measure annotations, our system's performance improves by $22\%$ relative (\macc{} $0.72 \to 0.88$)
on M13. However, we also observe a relative performance improvement of $14\%$ (\macc{} $0.72 \to 0.82$) using our system on the same dataset when we only pass in repeats. Given that repeats are much easier for humans to annotate than measure bounding boxes, 
we propose to have humans \emph{just label the repeats} as a recommended tradeoff between alignment quality and annotator time. 
We also explore 
providing our system with measure-level key signature and clef information, finding that this information
only marginally improves performance relative to the default key and clef assumptions described in~\Cref{sec:score-feature-rep}.

\vspace{-2.25mm}
\subsection{Different audio feature representations}

Here 
we compare alternative audio feature representations by evaluating \macc{} on M13 given all additional information (i.e.,~the R,M,S setting described in \Cref{sec:extra}). 
While we primarily use raw onset prediction probabilities (\Cref{sec:system-audio-feature-rep}),
the Onsets and Frames model \cite{hawthorne2017onsets} provides other possibilities including onset predictions (thresholded probabilities), 
frame probabilities and predictions, 
and the postprocessed MIDI transcription converted to piano roll (see \cite{hawthorne2017onsets} for details). Table \ref{tab:representation_exps} shows that onsets consistently outperform frames as an alignment representation---our intuition is that onsets are more appropriate for our setting as the bootleg score representation does not encode note durations. Additionally, while transcribed MIDI is a common feature representation for music alignment, 
in our setting we find it to be the worst choice.


\begin{table}[t]
 \begin{center}
 \begin{tabular}{lcc}
\toprule
Representation	& M13 & SMR\\
\midrule
Onset probabilities & 0.88 & 0.82 \\
Onset predictions & 0.86 & 0.82 \\
Frame probabilities & 0.70 & 0.53 \\
Frame predictions & 0.66 & 0.51 \\
MIDI & 0.46 & 0.20 \\

\bottomrule
 \end{tabular}
\end{center}
\caption{Evaluation of measure-aware alignment accuracies (\macc{}) achieved by different audio feature representations obtained from the Onsets and Frames piano transcription model \cite{hawthorne2017onsets} on MeSA-13 and SMR.}
 \label{tab:representation_exps}
\end{table}

\vspace{-2.5mm}

\vspace{-1.75mm}
\section{Labeling interface}
\label{sec:interface}

Here we describe a web-based interface that we built to enable experts to quickly annotate jumps in scores (induced by repeat signs, D.S. al coda, etc.). 
Our experiments show large improvements in alignment quality given jump labels. 
Accordingly, we designed an interface to make this process efficient---experts can label jumps in a matter of seconds. 
We include videos demonstrating the end-to-end process and qualitative results of our proposed workflow for pieces outside our evaluation data.\footnote{Video examples: \url{https://bit.ly/jltr-ismir2024}} 
In these videos, labeling jumps takes less than $6$s per page on average.

\begin{figure}[ht]
    \centering
    \includegraphics[alt={Illustration of our web-based labeling interface. The illustration contains a page from a score where individual measures are detected and can be clicked on in a specific order to indicate where the jumps occur in the page. The illustration also highlights features like hovering over detected measures and markers for clarifying where the jumps are annotated.},width=0.95\columnwidth]{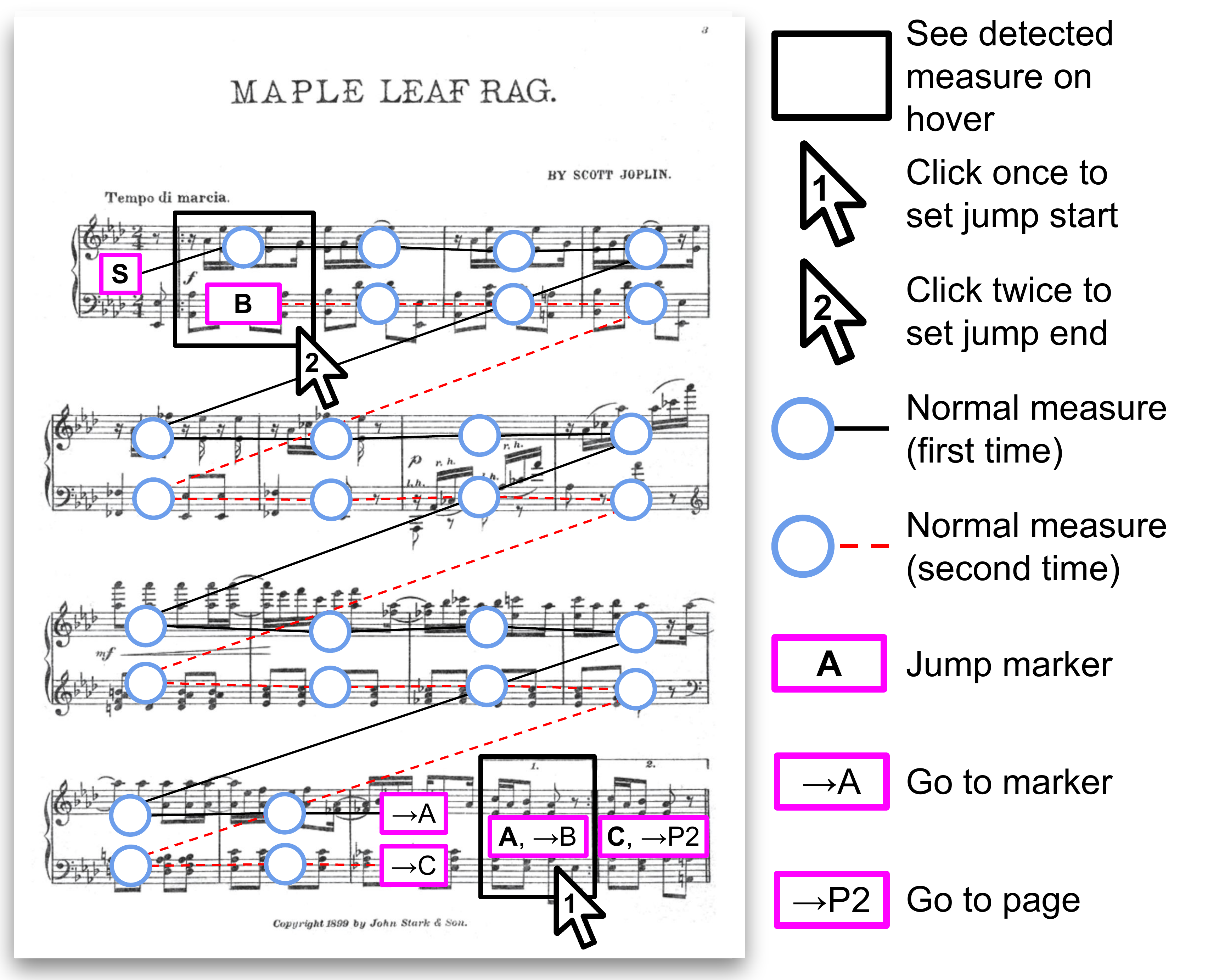}
    \caption{Illustration of our web-based interface for labeling jumps (e.g. repeats) in scores. 
    Our interface enables rapid jump annotation (just seconds per page after training), which we find to dramatically improve alignment quality on pieces with jumps.
    }
    \label{fig:interface}
\end{figure}

Our interface features a unified workflow for jump annotation based on clicking the starting and ending measure of a jump (\Cref{fig:interface}). 
To enable this workflow, we first run measure detection~\cite{waloschek2019identification} on the backend and visualize detected measures on the frontend as a user hovers over the score. 
Then, users can simply click on two different measures to set a jump---this simple unified workflow accommodates a long tail of jump glyphs.
The interface also visualizes the logical order of the measures induced by the measure bounding boxes and any jumps the user has set ($\mbb$ from~\Cref{sec:task_measure_aware}). 
Finally, a user can download the logical-order measures as a simple JSON file, which can then be loaded into our Python-based alignment package. 

\section{Related work}

Our work 
relates to prior work 
in offline audio-to-score 
alignment, 
score following, and 
score annotation tools.
\vspace{-2.25mm}
\subsection{Offline alignment}

\emph{Offline alignment} entails building a correspondence between sheet music and performance recording of the contained music.
As described previously, our work is inspired by the bootleg score line of research \cite{shan2020improved,shan2021automatic,tsai2020using,yang2020midi,yang2022large} as well as the usage of raw audio prediction scores in \cite{maman2022unaligned}. Our contributions stem from combining these approaches with minimal human annotations to outperform prior baselines. 
Other prior work in this area includes \cite{dorfer2017learning,dorfer2018learning}, which use 
MIDI events as an intermediate representation between scores and audio, and  \cite{kurth2007automated,damm2008multimodal,fremerey2009sheet,thomas2012linking,fremerey2010handling,gasser2013automatic}, which use 
chroma vectors (where components correspond to the 12 pitch types in Western music). 
We perform a similar procedure to the methods used in these works except that we use 
bootleg scores as our intermediate representation. We diverge from \cite{foscarin2020asap}, which uses MusicXML as an intermediate representation, and from \cite{liu2021joint}, which uses LilyPond representations, but we mention them here as related approaches.
We also incorporate reasoning from \cite{thickstun2020rethinking} regarding evaluation metrics. 

\vspace{-2.25mm}
\subsection{Score following}
In contrast to offline alignment, \emph{score following} involves building a real-time alignment between sheet music and live performance audio.
Initial solutions to this problem include \cite{dannenberg1984line,vercoe1984synthetic}, with later works addressing jumps and repeats \cite{pardo2005modeling, arzt2008automatic}; for more on related work in past decades, see survey papers \cite{puckette1992score,orio2003score}. Unlike most other research in score following which often assumes that a digital score representation (like MIDI) is available, our emphasis is on \emph{solely using score images} with an aim to perform alignment at scale.
This said, some recent work does attempt to solve this problem in images.
For instance, works such as \cite{dorfer2016towards,henkel2021real,henkel2020learning,henkel2022audio} map audio snippets to corresponding places in score images using neural networks, but they are limited to piano music. We diverge from them by considering a range of different types of raw score images and audio (such as ones with instrumentation beyond solo piano) and leveraging bootleg scores \cite{shan2020improved,shan2021automatic,tsai2020using,yang2020midi,yang2022large} for mapping, but these are still related to our work.
\vspace{-2.25mm}
\subsection{Sheet music annotation interfaces}
Our work also relates to past work on designing interfaces to assist in the annotation of sheet music.
Most directly related is that of Feffer et al.~\cite{feffer2022assistive} which attempts to facilitate interactive annotation of sheet music and audio alignment via a workflow based on aligning detected beat timestamps~\cite{madmom} to detected measures~\cite{waloschek2019identification}. 
This interface was used to compile the MeSA-13 dataset of aligned audio and scores, which we use to evaluate our work, though we note that their interface required $20$ hours of expert time to collect less than an hour 
of aligned data. 
In contrast, our interface is designed to be used for a matter of seconds to annotate repeats.
Other interfaces focus on facilitating measure bounding box annotations~\cite{zalkow2019tools,egozy2022computer}, which is complementary to our workflow that focuses on repeat annotation using predicted bounding boxes.
Lastly, Soundslice \cite{Soundslice} is a commercial product that offers 
an interactive alignment workflow based on stronger notions of OMR, but its implementation details are proprietary.
\vspace{-2mm}



\vspace{-1.75mm}
\section{Conclusion}
\vspace{-1mm}
In summary, we introduce a workflow for efficiently aligning sheet music images to performance audio. The key insight we leverage is that while automated alignment algorithms are currently not robust to repeats in scores, humans can quickly label these repeats, thereby improving alignment performance.
We validate this approach 
on a dataset of in-the-wild sheet music scans and real performance recordings, 
showing that we outperform existing baselines that only use automated approaches.

Given these results, one future project we aim to undertake is to collect jump annotations at scale to create large aligned datasets. We acknowledge that the datasets we used to evaluate our approach are small, but the insights gained from them can help scale up data for future evaluations.
For instance, we could extend the interface from Section \ref{sec:interface} to allow annotators to quickly audit and adjust alignments to collect more data, as in~\cite{feffer2022assistive}. Moreover, other future work could revisit the creation of a fully automated alignment algorithm with insights from our work, namely that such an algorithm that leverages OMR to identify jumps and repeats may be more successful than one that does not. Collecting more data would therefore be helpful for developing and evaluating future approaches. Lastly, as described in the start of our paper, large aligned datasets could be used to derive multimodal MIR systems for music students and professionals alike.

\section{Ethics Statement}








As described previously,
advancements in 
audio-to-score alignment can result in 
new multimodal datasets derived from existing repositories of sheet music and audio (such as IMSLP \cite{imslp}) 
and new interactive music systems tailored for performance. 
Our motivation for pursuing this direction is to unlock multimodal MIR systems
that 
(1)~supplement music education by helping performers rehearse, 
(2)~understand or generate sheet music to unlock seamless communication with human musicians, and
(3)~reduce reliance on copyrighted material for bulding music AI (i.e.,~by leveraging public domain scores and recordings).



We also recognize several potential ethical concerns stemming from our work. 
Firstly, our method is firmly rooted in conventions of Western music. Accordingly,  
downstream systems and data derived from our method may reflect a Western bias that does not generalize well to other musical traditions, especially those with different notation or tuning systems. 
Secondly, though our goal is to lessen the amounts of copyright infringement taking place to build generative music AI, 
multimodal MIR systems could be used to circumvent data protections, e.g., by transcribing copyrighted recordings as less-protected sheet music. 
Lastly, 
increased ability to understand sheet music could lead to deepfakes or misinformation, e.g., scores that could be falsely attributed to Beethoven, or ragtime recordings that could be falsely attributed to Joplin.
In response to these concerns, we recommend that future work mitigate these risks by, for example, developing analogous systems capable of improving understanding of non-Western music notation. We also recommend that 
MIR researchers
should be mindful of data protections, copyright violations, and artistic mimicry that, if subverted, could threaten the livelihood of musicians.


\bibliography{refs}

\end{document}